\documentclass[aps,prb,twocolumn,superscriptaddress,groupedaddress,longbibliography]{revtex4-1}
\usepackage{epsfig,amsopn}
\usepackage{graphicx}
\usepackage{wrapfig}
\usepackage{amsmath,amssymb}
\usepackage{enumerate}
\newcommand\bea{\begin{eqnarray}}
\newcommand\eea{\end{eqnarray}}
\newcommand\beq{\begin{equation}}
\newcommand\eeq{\end{equation}}

\def\nn{\nonumber}
\def\f{\frac}

\def\ga{\gamma}

\def\si{\sigma}
\def\Do{\partial}
\def\De{\Delta}

\def\ra{\rangle}
\def\ua{\uparrow}
\def\da{\downarrow}

\def\ka{\kappa}

\begin{document}
\title{Scattering in quantum wires and junctions of quantum wires with edge states of quantum spin Hall insulators}
\author{Abhiram Soori~~}
\email{abhirams@uohyd.ac.in}
\affiliation{ School of Physics, University of Hyderabad, C. R. Rao Road, 
Gachibowli, Hyderabad-500046, India.}

\begin{abstract} 
An integral part of scattering theory calculations in continuum quantum  systems involves identifying  appropriate  boundary conditions in addition to writing  down the correct Hamiltonian. In the simplest problem of scattering in one dimensional lattice, scattering due to an on-site potential and scattering due to an unequal bond (in otherwise translationally invariant lattice) give different results for scattering amplitudes. While the scattering problems in the continuum and the lattice models can be mapped to one another for scattering due to on-site potential, the equivalent continuum model for scattering due to an unequal bond is missing.  We introduce a new parameter $c$ in the boundary condition of the continuum  model that is equivalent to scattering due to an unequal bond on the lattice. Further,  we study a junction between a normal metal quantum wire and a one dimensional edge of quantum spin Hall insulator (QSHI) in continuum using the parameter $c$. In the case of a junction between a normal metal quantum wire and the edge of QSHI, we identify the boundary condition that permits maximum transmission. Further, we solve the scattering problem between the junction of quantum wire and QSHI using a lattice model and map it to continuum model results.  The problem of transport between four channels of spinful normal metal quantum wire and two channels of QSHI edge  is not well-defined. We rectify this situation by formulating the scattering problem in terms of a junction of a semi-infinite normal metal quantum wire with an infinite edge of QSHI, gapping out one semi-infinite section of the QSHI edge by a Zeeman field and applying an appropriate boundary condition at the junction. We calculate the scattering amplitudes analytically.  
\end{abstract}

\maketitle
\section{Introduction}
Scattering at a point in one dimensional continuum quantum mechanics is a
well studied text book problem, where a delta-function barrier back-scatters an 
electron~\cite{griffiths}. This phenomenon has been used in 
modeling point-like back-scatterers and interfaces extensively in  both 
noninteracting~\cite{soori10,amit07,btk,Pasanai16,suri2017} and interacting 
systems~\cite{kane92}. The point scatterers at the normal metal superconductor 
junction~\cite{btk} backscatters the electrons, reducing the probability of Andreev
reflection. In magnetic tunnel junctions~\cite{Pasanai16,suri2017}, the interface
modeled by delta function potential limits the electron transmission. 
In interacting one-dimensional systems~\cite{kane92}, 
repulsive short range interaction makes the  point like scatterer relevant 
in the sense of renormalization group and hence is  non-negligible. In this work, we 
restrict our discussion  to non-interacting electrons. While  the point  scatterers modeled by delta function barrier  in the continuum backscatter  electrons, on a lattice, a hopping
strength on a bond not equal to the other hopping strengths in an otherwise 
translationally invariant system (termed bond impurity) can also act as a scatterer. The scattering coefficients
for an on-site energy and for an unequal hopping strength on a lattice have different 
forms. The lattice model for the energy band can be mapped to the continuum model
near the band bottom. A natural question that arises then is- `what do the two types 
of scatterers in the lattice model map to when the lattice model is mapped to a 
continuum model?' We answer this question.

Topological insulators have attracted attention of researchers in the last decade
owing to their exotic properties such as dissipationless transport~\cite{hasan10,qi11}.
They are insulting in the bulk and conducting on the surface/edge. The first example of such
a phenomenon where insulating two-dimensional bulk accompanied by conducting edge states
dates back to quantum Hall effect~\cite{klitzing80}. It was later shown that conducting
edge states in a 
two-dimensional insulating bulk originating from the topology of the bulk band structure 
does not require a net magnetic field, and instead a staggered magnetic flux through a honeycomb
lattice can do the job~\cite{haldane88}. This was followed by a prediction by Kane and 
Mele that graphene with spin-orbit coupling is a two-dimensional topological 
insulator~\cite{kane05}. But spin-orbit coupling in graphene is too weak for the bulk gap 
to be observable. In 2006, Bernevig, Hughes and Zhang theoretically predicted that 
certain HgTe-CdTe quantum wells are two-dimensional topological insulators~\cite{Bernevig06}. 
Two-dimensional topological insulators also known as quantum spin Hall 
insulators~(QSHIs) are band insulators that have conducting one dimensional edge 
states which come in pairs. Soon after, in 2007 it was experimentally shown that in 
HgTe-CdTe quantum wells, QSHI can be realized~\cite{Konig07}. After a decade, a two-dimensional
material WTe$_2$ has also been shown to be a QSHI~\cite{jia17,Peng2017}. 

Junctions of topological insulators with normal metals are important, since such junctions 
are basic building blocks of electronic circuits involving topological elements. Junctions 
between two-dimensional surface states of three-dimensional  topological insulators with 
two-dimensional ferromagnets~\cite{modak12} and two-dimensional superconductors~\cite{soori13}
have been studied using a boundary condition which involves a new parameter $c$. Recently, 
this  boundary condition has been used in: explanation of planar Hall effect in topological 
insulators~\cite{suri21} and a proposal to identify Majorana bound states~\cite{lu21}. However,
the dependence of scattering amplitudes  on $c$ and the optimal value of $c$ which allows
maximum current across the junction is not known. The same boundary condition applies to a junction of one-dimensional normal metal with edge states of QSHI. 
In this work, we study the conductance across such a junction as a function of this parameter
and find the optimal value of the parameter $c$ for which the transmission is maximum. Further, we solve the scattering problem at a junction between a quantum wire and a QSHI using a lattice model and map the results to continuum model calculations. 
The problem of junction between two materials, each being semi-infinite, is common. But such a 
problem with one material being a spinful normal metal and the other - an edge state of QSHI is ill 
defined. In this work, we rectify this problem and construct such a junction. 

\section{Scattering in a normal metal quantum wire}
A point scatterer in continuum theory can be modeled either at
the level of Hamiltonian, where  the Hamiltonian has a Dirac delta
function in real space, or by a boundary condition in 
the wavefunction at the location of the barrier. Though these two
approaches are equivalent, we shall follow the latter approach since 
it can be more easily generalized. 
The Hamiltonian $H$, the wavefunction $\psi(x)$ 
and the boundary conditions can be written down for a metallic system
in the following way: 
\bea H ~=~ \Big( \frac{\hat{p}^2}{2m} -\mu \Big),&&~~\nn\\
\psi(0^-)=\psi(0^+),~~&&~~\Do_x \psi|_{0^-}^{0^+}= q_0 ~\psi(0),
\label{eq:ham1} \eea
where $\mu$~is the chemical potential which dictates the electron 
filling  of the quantum  wire, $\hat p$ is the momentum operator
and $q_0$ quantifies the strength of the  impurity which backscatters
the electron. 
The wavefunction for an electron incident from left to 
right at an energy $E$  takes the form: 
\bea 
\psi(x) &=& e^{i k x} + r_k ~e^{-i k x} {\rm ~~~for ~~}x < 0 \nn \\
       &=& t_k ~e^{i k x} {\rm ~~~for}~~~x \ge 0, ~\label{eq:psi1}
\eea
where $k=\sqrt{2m(\mu+E)}/\hbar$. 
By matching the boundary conditions, it can be shown that
\bea t_k={2ik}/{(2i k - q_0)} ~{\rm and}~
r_k={q_0}/{(2i k - q_0)} \label{eq:rktk-cont} \eea

This problem can be also stated on a one-dimensional lattice system
where on-site energy on one site is different from others in an 
otherwise regular lattice. The formula for scattering amplitudes is 
similar to eq.~\eqref{eq:rktk-cont} near the band bottom, where the
dispersion is quadratic. However, there is a 
different way of inducing backscattering on a regular infinite one
dimensional lattice, which is to simply change the hopping element
on one of the bonds in the otherwise translationally invariant
lattice. To define the problem more precisely, we 
resort to the second quantized language on an infinite one dimensional 
lattice:  
\bea 
H &=& -w \sum_{n\neq 0} ~[c^{\dagger}_{n+1} c_n + {\rm h.c.} ]
~~ -w' (c^{\dagger}_{1} c_0 + {\rm h.c.}) \nn \\ & & 
-(\mu-2w)\sum_n  c^{\dagger}_n c_n ,
\eea
where the hopping amplitude $w'$ may not be equal to $w$.
The scattering wavefunction for an electron incident from left to right has the form
\bea \psi_n &=& e^{ikan} + r_k e^{-ikan},~~~{\rm for~~}n\le 0,\nn\\
            &=& t_k e^{ikan}~~~{\rm for~~}n\ge 1, \label{eq:psi2}\eea
where $a$ is the lattice spacing and $ka=\cos^{-1}{[-(\mu-2w+E)/2w]}$. A scattering wavefunction corresponds to an electron extended throughout the system over an infinite number of sites, in contrast to a bound state whose wavefunction is confined to a particular region and decays exponentially away from the region~\cite{griffiths}.
When $w\neq w'$, a generic electron is backscattered.
Transmission  and reflection amplitudes can be found from Schr\"odinger 
equation to be
\bea 
t_k&=&\frac{-2iw' w\sin{ ka}  }{(w^2 e^{-ika}-w'^2 e^{ika})} \nn \\
 r_k&=&\frac{(w'^2-w^2)e^{ika}}{(w^2 e^{-ika}-w'^2 e^{ika})}~.
 \label{eq:rktk-latt}~~\eea

In the limit of small filling, the Fermi energy in both 
lattice and continuum models lies close to the band bottom.
Hence, the lattice dispersion can be approximated to a quadratic
dispersion, thereby mapping the continuum model to the lattice model. 
However, the expressions for the scattering amplitudes: eq.~\eqref{eq:rktk-cont}
and eq.~\eqref{eq:rktk-latt} cannot be mapped on to one another for small $k$. 
This means that the two kinds of scatterers are inequivalent.

To find the continuum equivalent of the unequal bond that causes backscattering, let us 
investigate the continuum theory. The boundary conditions given in eq.~\eqref{eq:ham1}
for the continuum theory come from continuity of the probability current on 
either sides of the impurity. This means $Im[\psi^*\Do_x\psi]$ is continuous at 
$x=0$. This implies that a more general boundary condition is
\bea \psi(0^-) &=& c\psi(0^+), {~~\rm and~~} \nn \\ 
\Do_x\psi|_{0^-} - q_{0^-}\psi(0^-) &=& \f{1}{c}[\Do_x\psi|_{0^+}
- q_{0^+}\psi(0^+)], \label{eq:bc-qw}\eea
where $c$, $q_{0^-}$ and $q_{0^+}$ are new parameters that take real values. $q_{0^-}$ ($q_{0^+}$)   physically means the strength of a delta function impurity at the location $x=0^-$ ($x=0^+$).  Now, solving for the 
scattering coefficients in the wavefunction given by eq.~\eqref{eq:psi1} from the above 
boundary condition, we get 
\bea 
t_k &=& \f{2ick}{ik(c^2+1)+c^2q_{0^-}-q_{0^+}}, {\rm ~~~and}~\nn \\
r_k &=& \f{ik(c^2-1)-c^2q_{0^-}+q_{0^+}}{ik(c^2+1)+c^2q_{0^-}-q_{0^+}}.
~\label{eq:rktk-cont2}\eea
In the limit of small $k$, eq.~\eqref{eq:rktk-latt} matches with eq.~\eqref{eq:rktk-cont2}
for the choice $c=w'/w$, $q_{0^-}=q_{0^+}=1/a$. Thus, we have  mapped the problem 
of unequal bond that causes backscattering to the continuum theory with appropriate boundary 
conditions. Physically, the new parameter $c$ in the boundary conditions corresponds to the extent 
to which the hopping parameter at the junction $w'$ is different from the hopping strength $w$
in the quantum wires. For the choice $q_{0^-}=q_{0^+}=1/a$, the limit
$c=1$ corresponds to the junction that allows perfect transmission, while a value of $c$ away from 
$1$ corresponds to an imperfect junction that results in backscattering. The boundary condition in eq.~\eqref{eq:bc-qw} does not follow from a continuum Hamiltonian. 

\section{Scattering at a junction of normal metal quantum wire and single edge of QSHI}
\begin{figure}[htb]
\begin{center}
 \includegraphics[width=8cm]{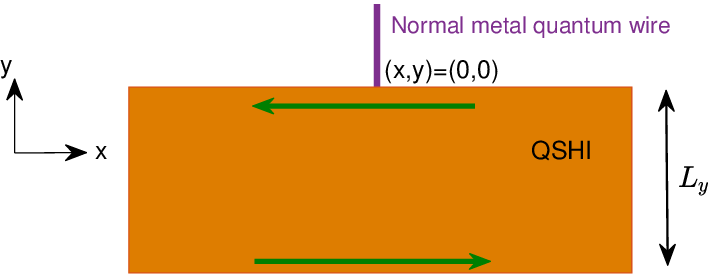}
\end{center}
 \caption{Schematic diagram of the junction proposed. Normal metal quantum wire (magenta) meets a QSHI. The  QSHI is taken to be infinite along $x$-direction, running from  $x=-\infty$ on the left extreme to $x=\infty$ on the right  extreme. The coordinate $y$ runs along the vertical direction with value $y=\infty$ on top
 extreme, a one dimensional  quantum wire extends from $y=\infty$ to $y=0$ (and $x=0$) to meet the QSHI at $(x,y)=(0,0)$. Green arrows show the up-spin edge states on the two boundaries of QSHI.}\label{fig-schem}
\end{figure}
\begin{figure}[htb]
 \includegraphics[width=8cm]{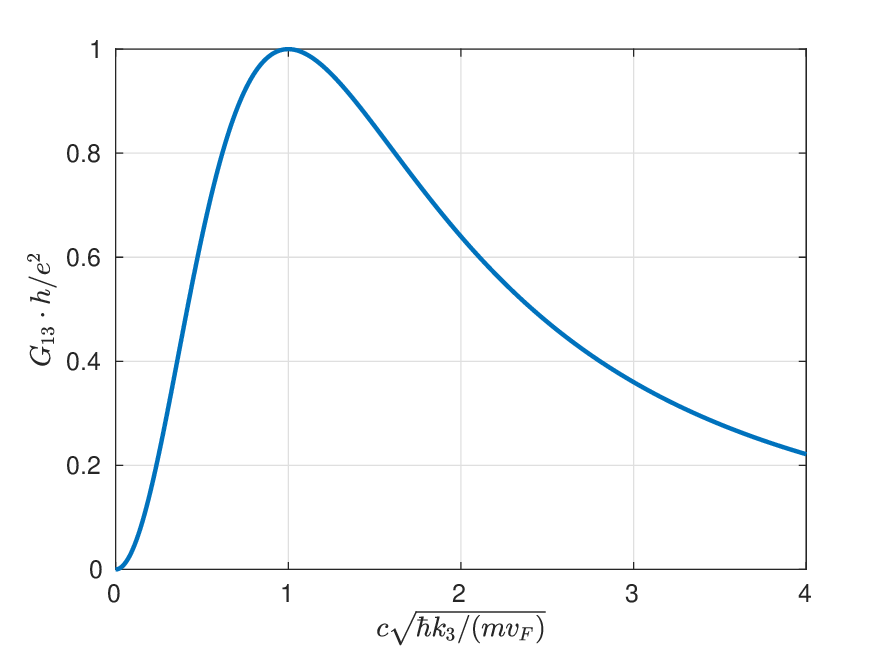}%-eps-converted-to.pdf}
 \caption{Conductance $G_{13}$ as a function of $c\sqrt{\hbar k_3/(mv_F)}$ from 
 eq.~\eqref{eq:g13}. It can be seen that when $c\sqrt{\hbar k_3/(mv_F)}=1$, the conductance
 is maximum and the transmission is perfect at $G_{13}=e^2/h$.}~\label{fig-gvsc}
\end{figure}

In this section, we study the case of the quantum wire being  connected to edge states at only one edge of the QSHI and the edge states on the two ends of the QSHI being decoupled. The Hamiltonian for edge states of QSHI is: 
\beq H = i\hbar v_F\si_z\Do_x, \eeq
where $\si_z$ is a Pauli spin matrix and $v_F$ is the Fermi velocity. 
Let us consider a normal metal quantum wire 
extending from $y=0$ to $y=\infty$ making a junction with QSHI edge at $x=0$, $y=0$. 
The schematic of the junction being considered can be seen in Fig.~\ref{fig-schem}.
The Hamiltonian for normal metal quantum wire is $H_3=(-\hbar^2\Do_y^2/2m-\mu)\si_0$, where
$m$ is the effective mass and $\mu$ is the chemical potential. We mark the three sides 
of the junction 1, 2 and 3, where side-1 corresponds to QSHI edge $x<0$, side-2 
corresponds to QSHI edge $x>0$  and side-3 corresponds to normal metal quantum wire 
$y>0$. We denote the wavefunctions in these three regions by $\psi_1$, $\psi_2$ and 
$\psi_3$ respectively. In regions 1 and 2, the spin is locked to the momentum. 
An up-spin electron with energy $E$ incident from the quantum wire onto the junction
has a wavefunction of the form:
\bea \psi_3(y) &=& e^{-ik_3y} |\ua\ra + r_{\ua} e^{ik_3y} |\ua\ra 
+ r_{\da} e^{ik_3y}|\da\ra,\nn \\
\psi_1(x) &=& t_1e^{-ik_xx}|\ua\ra, \nn \\
\psi_2(x) &=& t_2e^{ik_xx}|\da\ra,  \label{eq:psi123} \eea
where $|\ua\ra=[1,~0]^T$, $|\da\ra=[0,~1]^T$ are the spinors, 
$k_3=\sqrt{2m(\mu+E)}/\hbar$ and $k_x=E/(\hbar v_F)$. The time reversal invariant 
boundary condition that relates these wavefunctions is given 
by~\cite{modak12,soori13}:
\bea 
\psi_3 &=& c[M(\chi_1)\psi_1+M(\chi_2)\psi_2], \nn \\ 
\f{\hbar}{mv_F}\Do_y\psi_3-2\chi_3\psi_3 &=& -\f{i}{c}\si_z[M(\chi_1)\psi_1-M(\chi_2)\psi_2],~~ \label{eq:bc}
\eea
where $M(\chi)=\cos{\chi}-i\sin{\chi}\si_z$. This boundary condition conserves current. The parameters $\chi_i$, $i=1,2$
physically mean the barrier strengths on sides $i=1,2$ of the QSHI edge~\cite{sen12,*sen12err}.
Due to Klein tunneling, the barriers $\chi_1$ and $\chi_2$ on the QSHI edge allow perfect
transmission and
hence they can be set to zero. The calculation in the previous section suggests that the 
parameter $c$ is physically related to hopping from quantum wire to the QSHI edge. The 
parameter $\chi_3$ corresponds to the delta function barrier strength close to the junction on
the quantum wire. Let us set $\chi_i=0$ for $i=1,2,3$ and calculate the scattering coefficients in eq.~\eqref{eq:psi123} using the boundary condition in 
eq.~\eqref{eq:bc}. This gives us
\bea r_{\da} &=& t_2 = 0, \nn \\
t_1&=&\f{2\hbar k_3 c}{mv_F+c^2\hbar k_3}, \nn \\
r_{\ua} &=& \f{\hbar k_3 c^2-mv_F}{mv_F+c^2\hbar k_3}. \label{eq:rktk3} \eea
It is expected that the scattering coefficients in the spin down channels
are zero, since the incident electron is spin up and the full Hamiltonian
commutes with $\si_z$. 
The transmission amplitude $t_1$ is a function of $c$ and it can be shown that it is 
maximum for the choice $c=\pm\sqrt{mv_F/\hbar k_3}$. In fact, $c$ can be a function of 
energy and here, let us choose $c=\sqrt{mv_F/\hbar k_3}$. For this choice of $c$, 
$t_1=\sqrt{\hbar k_3/(mv_F)}$. This expression for transmission amplitude can have a 
value larger than 1. But the differential conductance $G_{13}=dI_1/dV_3$, 
the ratio of differential current on side-1 to the differential voltage applied on
side-3 is given by 
\beq G_{13} = \f{e^2v_F}{2\pi} \f{dk_3}{dE} |t_1|^2 
= \f{e^2}{h} \f{mv_F}{\hbar k_3}\Big[ \f{2c\hbar k_3}{mv_F+c^2\hbar k_3} \Big]^2, 
\label{eq:g13} \eeq
where the factor of $(1/2\pi)(dk_3/dE)$ is due to the density of states of the incident 
electrons. It can be easily shown that $G_{13}$ for the special choice of 
$c=\pm\sqrt{mv_F/(\hbar k_3)}$ will be $e^2/h$. In Fig.~\ref{fig-gvsc} we plot 
$G_{13}$ versus $c\sqrt{\hbar k_3/(mv_F)}$.

In other words, in an energy window between  $E$ and $E+dE$, the incident current on 
side-3 is:
\beq e \f{dk_3}{2\pi} v = e \f{dk_3}{2\pi}\f{dE}{\hbar dk_3} = \f{e}{h}dE \nn \eeq
The transmitted current on side-1 is $e D(E) dE|t_1|^2 v_F$ where $D(E)=(1/2\pi)(dk_3/dE)$
is the density of states of incident electrons. This is the same as $(e/2\pi)v_F|t_1|^2 dk_3$ 
which reduces to $(e/h)dE$ for the special choice of $c=\sqrt{mv_F/\hbar k_3}$.
Hence, though $t_1$ can have a magnitude larger than 1, the incident current is exactly 
equal to the  transmitted current due to the multiplicative factor $D(E)$. 

\section{Conductance of junction of quantum wire with QSHI from lattice model}
In this section, we describe transport calculation on a junction of quantum wire with QSHI described by tight binding model. QSHI is essentially two copies of Chern insulator, with one copy being the time reversed partner of the other~\cite{shen}. As we have seen in the previous section, the scattering problem can be solved separately in spin sectors that are eigenstates of $\si_z$ since $\si_z$ commutes with the full Hamiltonian. In this section, we study the scattering problem in the $|\ua\ra$-spin sector. One main qualitative difference  from the previous section is that, here, the width of the QSHI is finite and  the edge states from the two boundaries hybridize, opening up the possibility of an up-spin electron incident from the quantum wire getting transmitted either to the left or right on the QSHI. The lattice Hamiltonian for quantum wire can be written as $H_{3L}=\sum_{n=1}^{\infty}[-w(c^{\dag}_{n+1}c_n+{\rm h.c})-\mu_l c^{\dag}_nc_n]$, where $c_n$ is the annihilation operator for an electron at site $n$. The lattice for each spin component of QSHI is a bipartite lattice with two lattice sites per unit cell. Defining  $c_{n_x,n_y}=[c_{n_x,n_y,1},~c_{n_x,n_y,2}]^T$, where $c_{n_x,n_y,j}$ is annihilation operator at site $(n_x,n_y)$ with sublattice $j$, the Hamiltonian for QSHI can be written as
\bea  
H_Q &=& \sum_{n_x=-\infty}^{\infty}~ \sum_{n_y=-L_y+1}^{0}\Big[(\De-4B)c^{\dag}_{n_x,n_y}\tau_zc_{n_x,n_y}\nn \\ &&+c^{\dag}_{n_x+1,n_y}\big(B\tau_z+\f{iA}{2}\tau_x\big)c_{n_x,n_y}\nn \\ && +c^{\dag}_{n_x-1,n_y}\big(B\tau_z-\f{iA}{2}\tau_x\big)c_{n_x,n_y}\Big] + \sum_{n_x=-\infty}^{\infty}~ \sum_{n_y=-L_y+1}^{-1}  \nn \\ && \Big[ c^{\dag}_{n_x,n_y+1}\big(B\tau_z+\f{iA}{2}\tau_y\big)c_{n_x,n_y}+{\rm h.c.}\Big] ~,
\label{eq:H-tbm} \eea
where $\tau_i$ ($i=x,y,z$) are the Pauli spin matrices acting on the sublattice space. QSHI is in topological phase hosting edge states when $0<\De<8B$. 
The full Hamiltonian for the junction of quantum wire connected to a QSHI is $H=H_Q+H_{3L}+H_{WQ}$, where $H_{WQ}=-w_{nq}(c^{\dag}_{1}c_{0,0,1}+{\rm h.c.})$ and $w_{nq}$ is the hopping strength that connects the normal metal quantum wire to the QSHI.  

\begin{figure}
 \includegraphics[width=8cm]{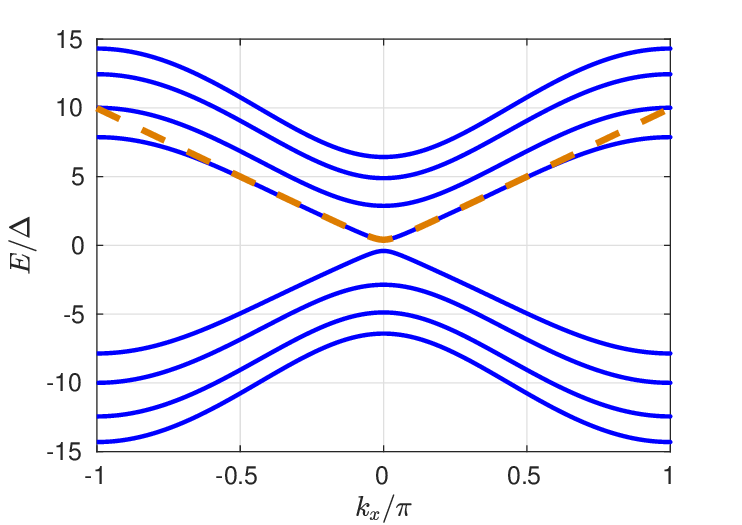}
\caption{Dispersion of QSHI lattice for $L_y=4$, $A=3\De$ and $B=2\De$. Yellow dashed line is a fit to the positive lowest energy band with the form $E=\sqrt{(\hbar v_Fk_x)^2+\ga^2}$. We get $\ga=0.4089\De$ and $\hbar v_F=3.171\De$.}\label{fig:qshi-disp}
 \end{figure}

\begin{widetext}
 
Using the above  Hamiltonian $H$ in Schr\"odinger wave equation, the eigenstate 
\bea |\psi\ra &=& \sum_{n=1}^{\infty}\psi^w_{3,n}|3,n\ra +\sum_{n_x=-\infty}^{\infty}\sum_{n_y=-L_y+1}^{0} \sum_{i=1,2}\psi_{n_x,n_y,i}|n_x,n_y,i\ra \label{eq:psi-l} 
\eea
(where $\psi^w_{3,n}$ is the wavefunction on the quantum wire and $\psi_{n_x,n_y,i}$ is wavefunction on the QSHI site) can be shown to obey the equations 
\bea
E\psi^w_{3,1}&=&-\mu_l\psi^w_{3,1}-w\psi^w_{3,2}-w_{nq}\psi_{0,0,1}, \nn \\
E\psi_{0,0}&=& -w_{nq}\begin{bmatrix} 
                       \psi^w_{3,1} \\ 0
                      \end{bmatrix} +
\big( B\tau_z+\f{iA}{2}\tau_y\big)\psi_{0,-1} + (\De-4B)\tau_z\psi_{0,0} +\big( B\tau_z+\f{iA}{2}\tau_x\big)\psi_{-1,0}  +\big( B\tau_z-\f{iA}{2}\tau_x\big)\psi_{1,0}, \nn  \eea \bea 
E\psi_{n_x,n_y}&=& \big( B\tau_z+\f{iA}{2}\tau_x\big)\psi_{n_x-1,n_y} +  (\De-4B)\tau_z\psi_{n_x,n_y}  +\big( B\tau_z-\f{iA}{2}\tau_x\big)\psi_{n_x+1,n_y} + \big( B\tau_z+\f{iA}{2}\tau_y\big)\psi_{n_x,n_y-1} \nn \\ && + \big( B\tau_z-\f{iA}{2}\tau_y\big)\psi_{n_x,n_y+1},  {\rm ~~ for~~}n_x=-1,0,1~{\rm and ~}  -L_y+2\le n_y \le-1, \nn \\ 
% E\psi_{0,n_y}&=& \big( B\tau_z+\f{iA}{2}\tau_x\big)\psi_{-1,n_y}+\big( B\tau_z-\f{iA}{2}\tau_x\big)\psi_{1,n_y} \nn \\ && + \big( B\tau_z+\f{iA}{2}\tau_y\big)\psi_{0,n_y-1} + \big( B\tau_z-\f{iA}{2}\tau_y\big)\psi_{0,n_y+1} \nn \\ && {\rm for~} -L_y+1\le n_y \le -1, \nn \\ 
E\psi_{n_x,-L_y+1}&=& \big( B\tau_z+\f{iA}{2}\tau_x\big)\psi_{n_x-1,-L_y+1} + (\De-4B)\tau_z\psi_{n_x,-L_y+1} +\big( B\tau_z-\f{iA}{2}\tau_x\big)\psi_{n_x+1,-L_y+1} \nn \\ && + \big( B\tau_z-\f{iA}{2}\tau_y\big)\psi_{n_x,-L_y+2},~~~ {\rm for ~~} n_x=-1,0,1 \nn \\
E\psi_{n_x,0}&=& \big( B\tau_z+\f{iA}{2}\tau_x\big)\psi_{n_x-1,0} + (\De-4B)\tau_z\psi_{n_x,0} +\big( B\tau_z-\f{iA}{2}\tau_x\big)\psi_{n_x+1,0}  + \big( B\tau_z+\f{iA}{2}\tau_y\big)\psi_{n_x,-1}, \nn \\ &&    {\rm ~~~for~}n_x=-1,1~,  \label{eq:eom}\eea
where $\psi_{n_x,n_y}=[\psi_{n_x,n_y,1},~\psi_{n_x,n_y,2}]^T$. The dispersion of quantum wire is $E=-2w\cos{k_3}-\mu_l$. The QSHI Hamiltonian $H_Q$ is translationally invariant along $x$ and has $2L_y$ bands. For $L_y=4$, $A=3\De$, and $B=2\De$, the dispersion for $H_Q$ is plotted in Fig.~\ref{fig:qshi-disp}. The bands closest to zero energy are the edge state bands. 

\end{widetext}

\begin{figure}
 \includegraphics[width=8cm]{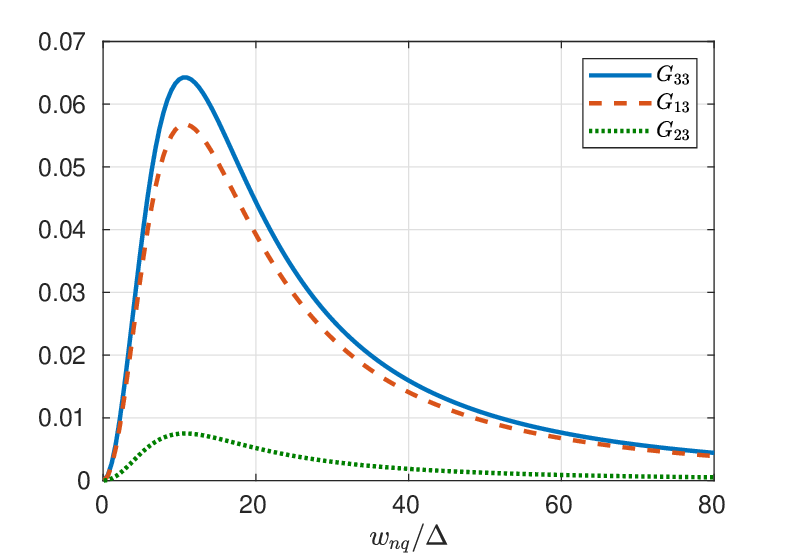}
\caption{ The conductances $G_{33},~G_{13},~G_{23}$ in units of $e^2/h$ versus $w_{nq}$ for $w=15\De$, $\mu_l=-30\De$, $A=3\De$, $B=2\De$ and $E=0.5\De$.}\label{fig:G-lattice}
\end{figure}

The scattering wavefunction for an electron incident from the quantum wire has the form:
\bea 
\psi^w_{3,n} &=& e^{-ikn} + r_k e^{ikn}, {\rm~~for~~} n \ge 1\nn \\ 
\psi_{n_x,n_y} &=& \sum_{j=1}^{2L_y} t_{R,j} e^{ik_{x,j}n_x} \chi_{n_y}(k_{x,j}), {\rm ~~for~~} n_x\ge 1, \nn \\ 
&=&  \sum_{j=1}^{2L_y} t_{L,j} e^{-ik_{k,j}n_x} \chi_{n_y}(-k_{k,j}), {\rm ~~for~~} n_x \le -1, \nn \\ && \label{eq:psi-lattice} 
\eea
where $\chi_{n_y}(k_x)=[\chi^{0}_{2n_y+1}(k_x),~\chi^{0}_{2n_y+2}(k_x)]^T$, $\chi^0(k_x)=[\chi^0_1,\chi^0_2,..,\chi^0_{2L_y}]^T$ is the eigenspinor of $H_Q$ with momentum $\hbar k_x$. For every $k_{x,j}$, $-k_{x,j}$ is also a solution of the dispersion relation. At a given energy $E$, $k_{x,j}$ is chosen so that if $k_{x,j}$ is real, $\partial E/\partial k_{x,j}>0$, and if $k_{x,j}$ is complex, ${\rm Im}[k_{x,j}]>0$. The values of $k_{x,j}$ at a given energy $E>\ga$ are found numerically. From eq.~\eqref{eq:eom}, the scattering coefficients $r_k$, $t_{L,j}$'s and $t_{R,j}$'s are found numerically for $w=15\De$, $\mu_l=-2w$ and  $w_{nq}$ in the range $[0.1,80]\De$. The currents in the QSHI for $n_x\ge 1$ (side-2) and $n_x\le 1$ (side-1) are carried by only the $k_{x,j_0}$ which is real and within the bulk gap, there is only one such pair $(k_{x,j_0},-k_{x,j_0})$. The differential conductance $G_{j,3}$ the ratio of infinitesimal change in current on side-$j$ to infinitesimal change in voltage on  side-3 (normal metal quantum wire) when the bias is changed from $E=eV$ to $E+dE=e(V+dV)$ is given by 
\bea 
G_{3,3}(E) &=& \f{e^2}{h}|r_k(E)|^2, \nn \\ 
G_{1,3}(E) &=& -\f{e^2}{h} \f{|t_{L,j_0}(E)|^2}{2w\sin k_3} \sum_{n_y=0}^{L_y-1}[\chi_{n_y}(-k_{x,j_0})]^{\dag}\nn \\ && \cdot [A\cos k_{x,j_0}\tau_x +2B\sin k_{x,j_0} \tau_z )]\chi_{n_y}(-k_{x,j_0}) , \nn \eea
\bea G_{2,3}(E) &=& \f{e^2}{h} \f{|t_{R,j_0}(E)|^2}{2w\sin k_3} \sum_{n_y=0}^{L_y-1}[\chi_{n_y}(k_{x,j_0})]^{\dag}[A\cos k_{x,j_0}\tau_x \nn \\ && -2B\sin k_{x,j_0} \tau_z )]\chi_{n_y}(k_{x,j_0}). \label{eq:cond}
\eea
 The conductances $G_{j3}$ are numerically evaluated using the lattice model and plotted in Fig.~\ref{fig:G-lattice} for $E=0.5\De$ as functions of the hopping amplitude $w_{nq}$. In the next section, we shall discuss a continuum model with boundary conditions that can be used to arrive at these results. 

\section{Continuum model calculations for a junction of quantum wire with QSHI edge states}

In this section, we propose the boundary conditions that characterize a junction between quantum wire and QSHI edge states. Similar to the discussion in the previous section, we shall focus our attention to $|\ua\ra$-spin sector. We shall map the lattice problem of the previous section to continuum. The Hamiltonian of quantum wire is $H_3=-\hbar^2\partial_y^2/2m$, where $m=\hbar^2/2w$ and $y$ takes values in the range $(0,\infty)$. The $|\ua\ra$-spin is left mover on the top edge and right mover on the bottom edge as depicted in Fig.~\ref{fig-schem} and the two modes are coupled. The Hamiltonian for $|\ua\ra$-spin edge states can be written as $H_Q=i\hbar v_F\eta_z\partial_x+\ga\eta_x$, where $\eta_x$ and $\eta_z$ are Pauli spin matrices whose components correspond to the top and the bottom edge, $\ga$ is the coupling strength that couples the two edges and $v_F$ is the Fermi velocity. The dispersion of QSHI edge states from this Hamiltonian is $E=\pm\sqrt{(\hbar v_F k_x)^2+\ga^2}$.   From the dispersion of QSHI lattice shown in Fig.~\ref{fig:qshi-disp}, values of $\ga$ and $\hbar v_F$ are found to be $\ga=0.4089\De$ and $\hbar v_F=3.171\De$ by mapping the continuum dispersion for small $k_x$. 

The current conservation at the junction between quantum wire and QSHI edge can be written as
\bea 
\f{\hbar^2}{m}{\rm Im}\Big[\psi_3^{\dag}\f{\partial\psi_3}{\partial y}\Big] &=& \hbar v_F[\psi_2^{\dag}\eta_z\psi_2 - \psi_1^{\dag}\eta_z\psi_1], 
\eea
all evaluated at $y=0$ and $x=0$. 
The boundary condition which satisfies current conservation is: 
\bea 
a\begin{bmatrix}
  p_1\psi_3 \\ q_1\psi_3
 \end{bmatrix} &=& c_1\psi_1, \label{eq:bc1} \\
 -a\begin{bmatrix}
   r_1\partial_y \psi_3 \\ s_1 \partial_y \psi_3 
  \end{bmatrix} &=& i\f{mv_F}{\hbar c_1}\eta_z\psi_1, \label{eq:bc2} \\ 
b\begin{bmatrix}
  p_2\psi_3 \\ q_2\psi_3
 \end{bmatrix} &=& c_2\psi_2, \label{eq:bc3} \\
 b\begin{bmatrix}
   r_2\partial_y \psi_3 \\ s_2 \partial_y \psi_3 
  \end{bmatrix} &=& i\f{mv_F}{\hbar c_2}\eta_z\psi_2, \label{eq:bc4}   
\eea
where $a,~b,~p_i,q_i,r_i,s_i$ are real-valued unknowns satisfying $a^2+b^2=1$ and $p_ir_i+q_is_i=1$ for $i=1,2$, $c_1$ and $c_2$ are variables used to define the boundary condition. Here, all the $\psi_i$'s and the derivatives are evaluated at $y=0$ or $x=0$. 
The scattering wavefunction has the form: 
\bea 
\psi_3(y) &=& e^{-ik_3y} + r_k e^{ik_3y}, \nn \\ 
\psi_1(x) &=& t_Le^{-ik_xx}[u_1,~v_1], ~~{\rm for~~}x<0,\nn \\ 
\psi_2(x) &=& t_Re^{ik_xx}[u_2,~v_2], ~~{\rm for~~}x>0, \label{eq:psi-cont}
\eea
where $k_3=\sqrt{2mE}/\hbar$, $k_x=\sqrt{E^2-\ga^2}/(\hbar v_F)$, $u_1=u_2=\ga$, $v_1=E-\hbar v_Fk_x$ and $v_2=E+\hbar v_Fk_x$.

\begin{figure}[htb]
 \includegraphics[width=8cm]{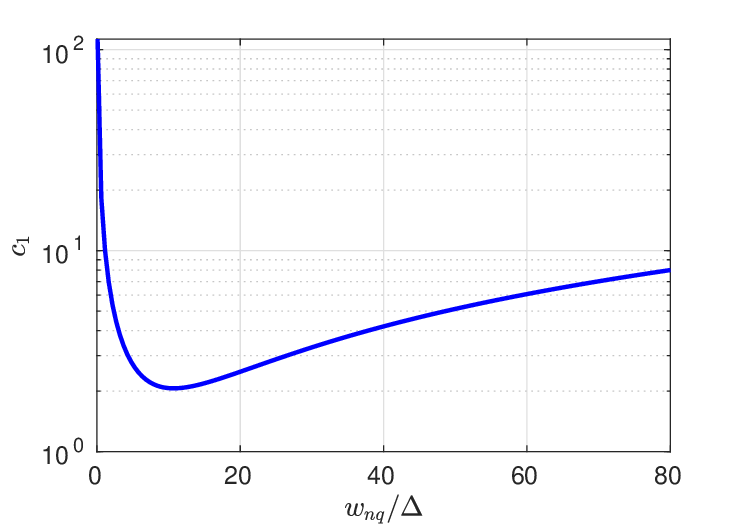}
 \caption{Value of $c_1$ obtained by fitting lattice model results of Fig.~\ref{fig:G-lattice} with continuum model result of eq.~\eqref{eq:rktk-cont} versus $w_{nq}$ for  $E=0.5\De$.}\label{fig:c1}
\end{figure}

Using the wavefunction in eq.~\eqref{eq:psi-cont} in the boundary condition equations -\eqref{eq:bc1}, \eqref{eq:bc2}, \eqref{eq:bc3}, and \eqref{eq:bc4}, further progress can be made. From eq.~\eqref{eq:bc1}, it can be seen that $p_1/q_1=u_1/v_1$ and  eq.~\eqref{eq:bc2} implies $r_1/s_1=-u_1/v_1$. Similarly, from eq.\eqref{eq:bc3} and eq.~\eqref{eq:bc4}, $p_2/q_2=u_2/v_2$ and $r_2/s_2=-u_2/v_2$. The unknowns $p_i, q_i, r_i, s_i$ can be parametrized by $\mu_i, \nu_i$ in the following way: $p_i=\mu_i u_i$, $q_i=\mu_i v_i$, $r_i=-\nu_iu_i$ and $s_i=\nu_iv_i$. The equation $p_ir_i+q_is_i=1$ implies $\mu_i\nu_i=1/(v_i^2-u_i^2)$. The unknown $\mu_1$ can be set to $1$. By finding the expression for $r_k$ from eq.~\eqref{eq:bc1} and eq.~\eqref{eq:bc2}, and equating that expression to the expression for $r_k$ obtained by solving  eq.~\eqref{eq:bc3} and eq.~\eqref{eq:bc4}, it can be shown that $\mu_2^2=-(c_1^2/c_2^2)[(v_1^2-u_1^2)/(v_2^2-u_2^2)]$. The parameters $a, b$ describe the fractions of the current from quantum wire that go into the left and the  right sides of QSHI and need to be specified. $a^2$ is the fraction of current incident from quantum wire that goes into side-1 of QSHI and the fraction $b^2$ goes into side-2.  The scattering coefficients $r_k, t_L, t_R$ can be found to be: 
\bea
r_k &=& \f{\hbar kc_1^2-mv_F(u_1^2-v_1^2)}{\hbar kc_1^2+mv_F(u_1^2-v_1^2)} \nn \\ 
t_L &=& -\f{2ac_1\hbar k}{\hbar kc_1^2+mv_F(u_1^2-v_1^2)} \nn \\ 
t_R &=& \f{2bc_1\hbar k}{\hbar kc_1^2+mv_F(u_1^2-v_1^2)} \sqrt{\f{u_1^2-v_1^2}{v_2^2-u_2^2}} \label{eq:rktk-cont3}
\eea
Here, while $c_1$, $a$ and $b$ will depend on $w_{nq}$, they may also may depend on energy $E$. The conductances can be found from these scattering coefficients by the formulae: 
\bea 
G_{33}(E) &=& \f{e^2}{h}(1-|r_k|^2) \nn \\ 
G_{13}(E) &=& \f{e^2}{h} \f{mv_F(u_1^2-v_1^2)}{\hbar k}|t_L|^2 \nn \\ 
G_{23}(E) &=& \f{e^2}{h} \f{mv_F(v_2^2-u_2^2)}{\hbar k}|t_R|^2 .
\label{eq:G-cont}\eea
From the data used to generate Fig.~\ref{fig:G-lattice}, by fitting the lattice model results for $G_{33}$, $r_k$ can be obtained using eq.~\eqref{eq:G-cont}. From this value of $r_k$, $c_1$ can be obtained by fitting with eq.~\eqref{eq:rktk-cont3}. The value of $c_1$ so obtained is plotted as a function of $w_{nq}$  in Fig.~\ref{fig:c1}. The parameter $a$ can also be fitted by fitting $G_{13}$ and at $E=0.5\De$, $a=0.9398$ for all values of $w_{nq}$. 

When the two edges of QSHI decouple, $u_1=1$ and $v_1=0$ and when $a=1$, the boundary condition in eq.~\eqref{eq:bc1}-\eqref{eq:bc4} reduces to the boundary conditions in eq.~\eqref{eq:bc} with $\chi_1=\chi_2=\chi_3=0$. 

\section{Junction of a normal metal quantum wire with semi-infinite QSHI edge}
The edge states of QSHI live on the boundary of a two-dimensional QSHI. Hence, they 
cannot be semi-infinite. But a Zeeman field perpendicular to the easy axis of the edge state 
electrons opens a gap in the spectrum of edge 
states~\cite{soori12} and this fact can be employed to make the infinitely long 
edge semi-infinite by applying a Zeeman field to a semi-infinite section of the 
QSHI edge. We achieve this by a modification of the system 
in the previous section where the Hamiltonians in regions 1 and 3 remain the same, while 
the Hamiltonian in region 2 is given by
\beq H_2= i\hbar v_F\si_z\Do_x +b \si_x. \eeq
We are interested in the energy range $|E|<b$. Also, we choose $\mu>b$. 
The wavefunction in the region 2 has the form 
\beq \psi_2(x) = t_2 e^{\ka x} |\ka\ra, ~\label{eq:psi-2}\eeq
where $\ka=\sqrt{b^2-E^2}/\hbar v_F$ and 
$|\ka\ra=[(E+i\hbar v_F\ka),~b]^T$ is the corresponding eigenspinor. 
We now solve the scattering problem for incident electrons from the quantum wire using
the boundary condition in eq.~\eqref{eq:bc} with the choice $\chi_i=0$ for $i=1,2,3$.

For an up spin electron incident from the quantum wire, the wavefunctions $\psi_1$ and $\psi_3$ have the same form as in eq.~\eqref{eq:psi123} and the scattering coefficients obtained on solving have the same form as in eq.~\eqref{eq:rktk3}. This is because, an up spin electron incident on the junction from the quantum wire has zero transmission amplitude in region-2 even in the limit of $b=0$.

For a down spin electron incident from the quantum wire, 
the wavefunction $\psi_1$  has the form as in eq.~\eqref{eq:psi123} and the
wavefunction $\psi_2$ has the form as in eq.~\eqref{eq:psi-2}. The wave function
in region 3 has a form
\beq \psi_3(y)= e^{-ik_3y} |\da\ra + r_{\ua} e^{ik_3y} |\ua\ra 
+ r_{\da} e^{ik_3y}|\da\ra, \label{eq:psi-3} \eeq
and the scattering amplitudes obtained after solving are
\bea
r_{\da} &=& \f{\hbar k_3c^2-mv_F}{\hbar k_3c^2+mv_F}, \nn \\
t_1 &=& \f{(E+i\hbar v_F \ka)}{b}\f{2\hbar k_3 c(mv_F-\hbar k_3c^2)}{(\hbar k_3c^2+mv_F)^2}, \nn \\
r_{\ua} &=& \f{(E+i\hbar v_F \ka)}{b}\f{4\hbar k_3 c^2mv_F}{(\hbar k_3c^2+mv_F)^2}.\label{eq:rktk3d}
\eea
Thus, we have shown how to construct a junction of a semi-infinite spinful quantum wire that has four channels (left moving and right moving channels for each of the two spins) with a semi-infinite  QSHI edge that has two channels and  given formulas for scattering coefficients for electrons incident onto the junction from the quantum wire. 

It is interesting to see that for $c=\sqrt{mv_F/\hbar k_3}$, 
the scattering coefficients  in eq.~\eqref{eq:rktk3d} reduce to $r_{\da}=t_1=0$, and 
$r_{\ua}=(E+i\hbar v_F \ka)/b$, which means that the electrons incident in the 
down spin channel get completely reflected into the up spin channel. This happens because of the
strong Zeeman field present in region 2 and in region 1 only up spin electrons can transmit. 
This means that a net spin current flows in region 3 due to a strong Zeeman field applied in region 2.

\section{Summary and Conclusion}
To summarize, we studied  scattering due to an on-site impurity in one dimension in continuum
and lattice models to show that the  scattering amplitudes in the two models can be mapped to each other.
Then we identified the boundary condition in the continuum theory that captures the problem of 
scattering due to a bond impurity by introducing a new parameter $c$. Thus, we have introduced a new boundary condition that describes a tunnel junction in a continuum model wherein the wavefunctions on either sides of the junction  may not be equal.  It may be noted that the wavefunctions (and their  derivative) on either sides of a junction  being equal does not conserve current at a junction between the spin orbit coupled region and a metal~\cite{soori21phesoc}. For the particle in a box problem, general boundary conditions wherein a linear combination of the wavefunction and its derivative is set to zero at an end of the box~\cite{carreau90,soori17} can be employed in place of the typical boundary condition which makes the wavefunction zero at the ends~\cite{griffiths}. Similarly, the boundary condition introduced in this work for scattering across a point like impurity in a quantum wire is more general.

We also studied the problem of a junction of a quantum wire with infinite edge of QSHI using a boundary condition that uses the parameter $c$. 
If the junction between quantum wire and QSHI is described by a lattice model, it is the hopping from quantum wire to the QSHI  that is responsible for electron transport from quantum  wire to the QSHI edge. This suggests that the parameter $c$ is related to hopping strength between quantum  wire and QSHI. We investigated the dependence of the scattering amplitudes on $c$ and find  the value of $c$ for which the transmission across the junction is perfect. A realistic QSHI will have two edges and the parameters $c$ will have two values $c_1$ and $c_2$ corresponding to the couplings to the two edges. We  perform a systematic study starting from  a lattice model for a junction between quantum wire and QSHI and find the parameter $c_1$ as a function of the hopping amplitude $w_{nq}$ that connects the quantum wire to QSHI. 
Further, we construct a  model in which four channels of spinful quantum wire are connected  with two channels of QSHI edge by introducing a Zeeman field over a semi-infinite patch on QSHI edge in the direction perpendicular to the spin easy axis of the edge state electrons. If $\hat z$ is the direction of spin of the electrons moving away from the junction in the QSHI edge in such a junction, the incident  electrons from quantum wire  with spin 
pointing along $-\hat z$ flip their spin into $\hat z$ direction completely on reflection and the transmission probability onto the QSHI edge state is zero for a special value of $c$. This fact could be useful in spintronic applications. 

A junction of quantum wire with QSHI edge states that transmits perfectly at all energies is 
characterized by the boundary condition in eq.~\eqref{eq:bc} with the choice of parameters 
$\chi_3=0$ and $c=\sqrt{mv_F/\hbar k_3}$. This means $c$ needs to be a function of energy. The exact dependence of $c$ on energy in the generic system needs to be derived from the lattice model of the full system. However,  it can be said that $c$ is a smoothly varying function of energy and at some particular energy at which $c=\sqrt{mv_F/\hbar k_3}$, the transmission is perfect. 

It has been proposed that edge states of QSHI host exotic Majorana fermions at the  interface of a ferromagnet (the region exposed to Zeeman field) and a region that has superconductivity~\cite{fu09}. Zero bias conductance peak is a feature of Majorana fermions.  Such a junction can possibly be probed by an external STM tip for zero bias conductance peak~\cite{lu21}.  The boundary condition discussed in this work can be used to describe the STM tip connected to a ferromagnet-superconductor junction on QSHI edge. In future devices which consist of QSHI and normal metal quantum wires as components, the boundary conditions described in this work can be used to calculate transport properties. 

\acknowledgements
The author thanks  Deepak Dhar, Sumathi Rao and Diptiman Sen for stimulating discussions and Ranjan Laha for help with Mathematica. The author also thanks anonymous referee for useful comments. The author thanks DST-INSPIRE Faculty Award (Faculty Reg. No.~:~IFA17-PH190) for financial support.
\bibliography{ref-tinm}

\end{document}